\documentclass[aps,prl,twocolumn,superscriptaddress]{revtex4}
\usepackage{amsmath}
\usepackage{graphicx}
\usepackage{dcolumn}
\usepackage{bm}

\makeatletter
\def\captionof#1#2{{\def\@captype{#1}#2}}
\makeatother
\newcommand{\rme}{{\rm e}}
\newcommand{\rmi}{{\rm i}}
\newcommand{\ud}[1]{\hspace{-0em}\mathrm{d}{#1}\;}
\newcommand{\udd}[2]{\hspace{-0em}\mathrm{d}{#1}\,\mathrm{d}{#2}\;}
\newcommand{\Ud}[1]{\hspace{-0.5ex}\mathrm{d}{#1}\;}

\begin{document}

\title{Dimensional Crossover Driven by an Electric Field}

\author{Camille Aron}
\author{Gabriel Kotliar}
\affiliation{Department of Physics and Astronomy, Rutgers University,
136 Frelinghuysen Rd., Piscataway, NJ 08854, USA}
\author{Cedric Weber}
\affiliation{Cavendish Laboratories, Cambridge University,
JJ Thomson Avenue, Cambridge, UK}

\begin{abstract}
 We study the steady-state dynamics of the Hubbard model driven out-of-equilibrium by a constant electric field and coupled to a dissipative heat bath.
 For very strong field, we find a dimensional reduction: the system behaves as an \textit{equilibrium} Hubbard model in lower dimensions.
 We derive steady-state equations for the dynamical mean-field theory in the presence of dissipation.
 We discuss how the electric field induced dimensional crossover affects the momentum resolved and integrated spectral functions, the energy distribution function, as well as the steady current in the non-linear regime.
\end{abstract}

\maketitle
There is a growing interest in quantum physics out of equilibrium motivated by new experimental realizations in non-linear transport in devices, heterostructures and cold atoms~\cite{nano, BlochOscOptics,cold-atoms}.
On the theoretical side, new techniques are being developed in the context of correlated electrons driven by electric field, ranging from exact diagonalization and density matrix renormalization group techniques for treating one-dimensional systems to extensions of the dynamical mean-field theory (DMFT) in out-of-equilibrium situations~\cite{Oka2003,neqDMFT, Freericks2006,EcksteinWerner2010}. Important problems such as the suppression of the Bloch oscillations by the electronic interactions~\cite{Freericks2008} and the dielectric breakdown of the Mott insulator~\cite{EcksteinOkaWerner2010} have been studied with these methods.

In this paper, we focus on the non-equilibrium steady-state dynamics of a correlated metal driven by a static and uniform electric field.
We formulate a consistent description of the steady-state by including a thermostat, which plays a key role to prevent the subsequent current to heat up the system.
We show analytically that the system undergoes a dimensional crossover: for very strong fields,
the physics reach thermal equilibrium in lower dimensions. We derive steady-state equations for the DMFT in this driven scenario and simply solve the impurity by the so-called iterated perturbation theory (IPT). We study the effects of the electric field and illustrate the dimensional crossover by computing experimentally accessible quantities including the spectral function, the energy distribution function or the steady current.

\paragraph{Model.}
We consider the $d$-dimensional Hubbard model on a hypercubic lattice of lattice spacing $a$. A static and uniform electric field is set along one of the axes of the lattice: $\mathbf{E} = E \mathbf{u}_x$ with $E>0$.
The Lagrangian of the system is given by (we set $\hbar=1$)
\vspace{-0.01em}
\begin{equation}
\begin{array}{rl}
\mathcal{L}_s =& \displaystyle \sum_{i \sigma} \bar c_{i \sigma} \left[ \rmi\partial_t -  \phi_{i}(t) \right] c_{i \sigma} 
 - U \sum_{i}  \bar c_{i \uparrow}  c_{i  \uparrow}  \bar c_{i \downarrow}  c_{i \downarrow} \\ 
  & +  \displaystyle \sum_{\langle i j \rangle \sigma}   \bar c_{i \sigma} t_{ij} \rme^{\rmi\alpha_{ij}(t)} c_{j \sigma} 
+ \mathrm{conj.}\,,
\end{array}
\end{equation}
where $c_{i\sigma}$ and $\bar c_{i\sigma}$ are the Grasmann fields representing an electron at site $i$ with spin $\sigma\in\{\uparrow,\downarrow\}$. $U$ is the on-site Coulombic repulsion between electrons and $t_{ij} \equiv (a/2\pi)^{d} \int\Ud{\mathbf{k}} \rme^{\rmi\mathbf{k}\cdot\mathbf{x}_{ij}} \epsilon(\mathbf{k}) $ sets the hopping amplitude between two neighboring sites. Because of the periodicity in the lattice, all integrals in $\mathbf{k}$-space are computed in the Brillouin zone (\textit{i.e.} between $-\pi/a$ and $\pi/a$ in each direction). 
The Peierls phase factors $\alpha_{ij}(t) \equiv q \int_{\mathbf{x}_j}^{\mathbf{x}_i} \ud{\mathbf{x}} \cdot \mathbf{A}(t,\mathbf{x})$ are required by the gauged $\mathrm{U}(1)$ symmetry associated with the conservation of the charge $q$ of the electrons.
We gather the scalar potential $\phi$ and the vector potential $\mathbf{A}$  in the gauge field $A^\mu\equiv(\phi/q,\mathbf{A})$.

The electric field acts as a source of energy and we provide a heat sink by coupling each site $i$ to an independent thermal bath~\cite{Naoto, Adriano}.
These are reservoirs of non-interacting electrons in equilibrium (and not affected by the electric field) at temperature $T$ and chemical potential $\mu_0$. The latter controls the electronic occupation of the system.
We choose a gauge-invariant bi-linear coupling with the system: $\mathcal{L}_{sb} = \gamma \sum_{i \sigma l} \rme^{\rmi\theta_i(t)} \bar b_{i\sigma l} c_{i\sigma} + \mbox{conj.} $ where $\gamma$ is a coupling constant, the $b$'s represent the bath electrons, $l$ labels their energy levels and $\theta_i(t) \equiv  \int^t \Ud{t'} \phi_{i}(t')$. A regular dissipation in the whole spectrum is achieved if the bandwidth $W$ of the reservoirs is larger than any other energy scale.

The energy scales involved are the strength of the hopping controlled by $\epsilon_0\equiv\epsilon(\mathbf{k}=0)/\sqrt{2d}$, the strength of the electronic repulsion $U$, the temperature $T$ that sets the scale of the thermal excitations  (we set $k_\mathrm{B}=1$), and the strength of the dissipation $\Gamma \equiv \gamma^2/W$. The energy scale introduced by the electric field is given by $|q|Ea$.

\paragraph{Steady-state equations.---}
We analyze the dynamics by use of the Schwinger-Keldysh formalism~\cite{Keldysh}. Correlations and field being switched on far in the past, the properties of the steady state do not depend on the details of the initial conditions.
The steady-state physics is invariant in space and time. In particular, a mean-field analysis should not depend on the location and time at which the impurity is singled out from the bulk. This motivates us to work with a covariant formalism
rather than imposing a particular gauge which would artificially break the symmetry of the problem.
Steady-state equations are then derived --without having to solve the transient dynamics-- by imposing space and time translational invariance.

To shorten the notations, we drop the spin indices and, while working on the lattice, we adopt continuous space notation~\cite{Buot}: 
space-time coordinates are given by  $x^\mu\equiv(t,\mathbf{x})$. The Schwinger-Dyson equations in the Keldysh basis read:
\begin{align*}
  &  D(x,x_1) \circ_{x_1} \!G^{R}(x_1,x') = \delta(x - x')\,, \\
  & G^K(x,x') =  G^{R}(x,x_1) \circ_{x_1} \! \Sigma^K(x_1,x_2) \circ_{x_2} \! G^{R}(x', x_2)^*\,,
\end{align*}
with $ D(x,x') \equiv  \delta(x-x')  \left[  \rmi\partial_{t'} - \phi(x)  \right] + \delta(t-t') t(\mathbf{x}-\mathbf{x}') \rme^{\rmi \alpha(x,x')}  - \Sigma^R(x,x')$ and $\circ_{x_1}$ is the convolution product (inner summation on $x_1$). $G^R$ and $G^K$ are respectively the retarded and the Keldysh components of the 2 by 2 Green's function matrix: $G^{R} \equiv G^{++}-G^{+-}$ and $G^K \equiv \rmi [G^{+-}+G^{-+}]/2$.
Both the thermal environment and the electronic repulsion contribute to the self-energy: $\Sigma^{R/K}\equiv\Sigma_{th}^{R/K}+\Sigma_{U}^{R/K}$.
In equilibrium, Green's functions (and self-energies) are related through the fermionic fluctuation-dissipation theorem (FDT): $G^K(k) = - \tanh\left(\frac{\omega-\mu_0}{2T}\right) \mbox{Im}\, G^R(k)$ (the same relation holds between $\Sigma^K$ and $\mbox{Im}\, \Sigma^R$) where $k^\mu \equiv (\omega,\mathbf{k})$ are the Fourier conjugates of $x^\mu$.

After a Wigner transform~\footnote{A Wigner transform of a two-point function is a Fourier transform in the relative coordinates: $f(X;k) \equiv \int\ud{x} \rme^{\rmi k_\mu x^\mu} f\left(X+\frac{1}{2}x,X-\frac{1}{2}x\right)$.}, the Schwinger-Dyson equations are expressed in terms of the center-of-mass coordinates $X^\mu$
and the conjugated variables $k^\mu$ of the relative coordinates  $x^\mu$:
\begin{align*}
   & \displaystyle D(X;k) \star G^{R}(X;k) = 1\,, \\
   & \displaystyle G^K(X;k) =  G^{R}(X;k) \star \Sigma^K(X;k) \star G^{R}(X;k)^*\,.
\end{align*}
The star product is the non-commutative Moyal product defined by $\star\equiv\exp{(\frac{\rmi}{2} [ \overleftarrow{\partial_{k^\mu}} \overrightarrow{\partial_{X_\mu}} \!- \overleftarrow{\partial_{X^\mu}} \overrightarrow{\partial_{k_\mu}} ])}$~\cite{Moyal}. Left (resp. right) arrows indicate that the derivative operators act on the left (resp. right) and the Einstein summation convention is used.

Schwinger-Dyson equations can be rendered explicitly covariant using the covariant derivatives $\rmi\partial_\mu - q A_\mu(X)$ and their Wigner transforms $\kappa^\mu(k,X) \equiv k^\mu - q A^\mu(X) \equiv (\varpi,\boldsymbol{\kappa})$~\cite{Onoda}.  
Once Green's functions are re-expressed in terms of these new variables, $\tilde G(X;\kappa)\equiv G(X;\kappa+qA)$, the star product becomes 
$
\exp{(\frac{\rmi}{2}
[
 \overleftarrow{\partial_{\kappa^\mu}} \overrightarrow{\partial_{X_\mu}}\!- \overleftarrow{\partial_{X^\mu}} \overrightarrow{\partial_{\kappa_\mu}}
\!+  q\overleftarrow{\partial_\varpi} \overrightarrow{\mathbf{E} \cdot \boldsymbol{\nabla}_{\!\boldsymbol{\kappa}}}
\!-  q\overleftarrow{\mathbf{E} \cdot \boldsymbol{\nabla}_{\!\boldsymbol{\kappa}}} \overrightarrow{\partial_\varpi}  
])
} 
$. In the rest of the manuscript, we simplify the notations by dropping the tilde.

For uniform and stationary solutions, the dependence on $X$ is lost and the equations reduce to
\begin{align}
 & D(\kappa) \ast  G^R(\kappa) = 1\,, \label{eq:StSt:GR} \\
 & G^K(\kappa) = G^R(\kappa) \ast \Sigma^K(\kappa) \ast G^R(\kappa)^*\,, \label{eq:StSt:GK}
\end{align}
with $D(\kappa) = \varpi + \epsilon(\boldsymbol{\kappa}) - \Sigma^R(\kappa)$ and the star product is simplified to  $\ast\equiv 
\exp{(\frac{\rmi}{2}q
[ 
 \overleftarrow{\partial_\varpi} \overrightarrow{\mathbf{E} \cdot \boldsymbol{\nabla}_{\!\boldsymbol{\kappa}}}
\!- \overleftarrow{\mathbf{E} \cdot \boldsymbol{\nabla}_{\!\boldsymbol{\kappa}}} \overrightarrow{\partial_\varpi}  
])
}$. In the $E=0$ limit, the star product between Wigner transforms reduces to the usual product between Fourier transforms and the equilibrium equations are immediately recovered. The quantum Boltzmann transport theory~\cite{Mahan} corresponds to performing a gradient expansion in $\mathbf{E}$.

\paragraph{Dimensional reduction.} 
When the energy scale associated to the electric field ($|q|Ea$) is much larger than any other relevant energy scale, the dimensionality of the system is reduced since the hoppings along the direction of the field are inhibited as electrons experience Bloch oscillations with a vanishing amplitude $\epsilon_0/|q|E$ (see \textit{e.g.}~\cite{Wilkins} for the non-interacting case). These turn into a steady state in the presence of a small scattering.
The very strong field limit can be worked out analytically in Eqs.~(\ref{eq:StSt:GR}) and (\ref{eq:StSt:GK}), showing that Green's functions (and therefore self-energies as well) lose their dependence along the direction of the field (\textit{i.e.} in $\kappa_x$) and the problem reduces to the \textit{equilibrium} Hubbard problem in \textit{lower} dimensions~\cite{detailed}:
\begin{align}
  G^R(\kappa)  \to& 
\left[ \varpi + \epsilon( \boldsymbol{\kappa_\perp}) -\Sigma^R(\varpi,\boldsymbol{\kappa}_\perp) \right]^{-1}\,, \\
 G^K(\kappa)  \to& 
\left|   
G^R\left( \varpi , \boldsymbol{\kappa}_\perp  \right) 
\right|^2  \Sigma^K(\varpi, \boldsymbol{\kappa}_\perp) \,,
\end{align}
with $\epsilon(\boldsymbol{\kappa}_\perp) \equiv
(a/2\pi) \int\ud{\kappa_x}\epsilon(\boldsymbol{\kappa}_\perp+\kappa_x\mathbf{u}_x)$ the corresponding lattice dispersion relation. 
This crossover from an out-of-equilibrium problem in $d$ dimensions to an equilibrium problem in lower dimensions constitutes one of the main results of this work.
When the field has sizeable projections on several principal axes, those directions are similarly suppressed.
It extends to any model on a periodic lattice under the assumption that a steady state can be reached and it is valid prior to any approximation.
Below, we support this result by following the crossover of selected observables in a two-dimensional system.
We obtain the results by a mean-field approximation for $\Sigma_U$ which we present now.

\paragraph{Dynamical-Mean Field Theory.---}
The DMFT is a non-perturbative approximation scheme, particularly well suited for strongly correlated systems in equilibrium.
Here, we extend it to our non-equilibrium steady-state regime. 
The self-energy $\Sigma_U(\kappa)$ arising from the electron interactions is taken to be local (\textit{i.e} independent of $\boldsymbol{\kappa}$, which is exact in the $d\to\infty$ limit) and equal to the one of an impurity problem associated self-consistently to the original lattice problem~\cite{DMFT}.
The action of the impurity reads  in the $(a,b\in\pm)$-basis
\vspace{-0.5em}
\begin{equation}
 \begin{array}{lcl}
S&=&\displaystyle\sum_{ab} \sum_{\sigma} \iint\udd{t}{t'} \bar c_\sigma^a(t) {\mathcal{G}^{-1}}^{ab}(t-t') c_{\sigma}^b(t') \\ & &\displaystyle - \sum_{a} a \int \ud{t} U    \bar c^a_{\uparrow}(t) c^a_{\uparrow}(t)  \bar c^a_{\downarrow}(t)  c^a_{\downarrow}(t)\,,
\end{array}
\vspace{-0.5em}
\end{equation}
where the impurity non-interacting  Green's functions (often referred as the Weiss effective fields) are determined self-consistently through
\begin{align}
 \mathcal{G}^R(\varpi) =& \left[ G^R_\mathrm{}(\varpi)^{-1} + \Sigma_U^R(\varpi) \right]^{-1}  \,,
\label{eq:WeissR}\\
\mathcal{G}^K(\varpi)
=& \left| \mathcal{G}^R(\varpi)\right|^2 \left[ \frac{ G^K_\mathrm{}(\varpi)}{\left|  G^R_\mathrm{}(\varpi)\right|^2 } - \Sigma_U^K(\varpi) \right] \,,\label{eq:WeissK}
\end{align}
with $G_\mathrm{}^{R/K}(\varpi) \equiv (a/2\pi)^d \int \Ud{\boldsymbol{\kappa}} G^{R/K}(\kappa)$.

Let us recast Eq.~(\ref{eq:StSt:GR}) into
\begin{align}\label{eq:DMFT:GR}
 G^R(\kappa) = G^R_0(\kappa) + G^R_0(\kappa)\ast \Delta\Sigma^R(\varpi) \ast G^R(\kappa)\,,
\end{align}
where $G^R_0(\kappa)$ is the non-interacting solution of $G^R_0(\kappa) \ast D_0(\kappa)=1$ with $D_0(\kappa) \equiv \varpi+\epsilon(\boldsymbol{\kappa}) - \Sigma^R_0$ and $\Delta\Sigma^R(\varpi) \equiv\Sigma^R(\varpi)-\Sigma^R_0$.
$\Sigma^{R}_0$ is an arbitrary `dissipative' constant ($\mbox{Im}\, \Sigma^{R}_0 <0$) which helps the convergence of the integrals below. 
Once functions are expressed in the mixed $(\tau;\boldsymbol{\kappa})$-space (after a Fourier transform to the real-time domain) one has $G^R_0(\tau;\boldsymbol{\kappa}) =  -\rmi
 \exp\left[{{\rmi}\int_{-\tau/2}^{\tau/2} \Ud{\tau'}   \epsilon\left( \boldsymbol{\kappa} + {q\mathbf{E}} \tau' \right) -{\rmi} \tau \Sigma^R_0 } \right] \Theta(\tau)
$ where $\Theta(\tau)$ is the Heaviside step function.
Equations~(\ref{eq:WeissR}) and (\ref{eq:WeissK}) together with Eqs.~(\ref{eq:StSt:GK}) and (\ref{eq:DMFT:GR}) generalize the DMFT to our driven steady-state scenario and constitute one of the main results of this work.

Our DMFT algorithm in the steady state goes along the following lines. 
(i) The lattice Green's functions are determined.
$G^R$ is obtained by solving Eq.~(\ref{eq:DMFT:GR}) self-consistently starting from $G^R\!:=G^R_0$ and from a guess for the impurity self-energy (typically $\Sigma^{ab}_U:=0$). The right-hand-side of Eq.~(\ref{eq:DMFT:GR}) is evaluated numerically \textit{via} a discretized form of the following integral representation of the star product:
\vspace{-0.5em}
\begin{equation}\label{eq:MoyalP}
 \begin{array}{rl}
 \left[ f \ast g\right] (\tau;\boldsymbol{\kappa}) =& {\displaystyle \int} \textstyle \ud{\tau'} f\left(\tau-\tau';\boldsymbol{\kappa}+{q\mathbf{E}}\frac{\tau'}{2}\right)   \\ 
 &  \quad \textstyle \times g\left(\tau';\boldsymbol{\kappa}+{q\mathbf{E}}\frac{\tau'-\tau}{2} \right)\,.
\end{array}
\vspace{-0.5em}
\end{equation}
 $G^K$ is then obtained similarly from Eq.~(\ref{eq:StSt:GK}).
 (ii) The Weiss fields of the impurity are determined by Eqs.~(\ref{eq:WeissR}) and (\ref{eq:WeissK}).
(iii) The self-energy $\Sigma_U$ is then obtained by solving the impurity problem in the steady-state. In principle this can be done by any out-of-equilibrium method. We simply solve it by means of IPT, treating the on-site interaction to the second order in $U$:
\vspace{-0.5em}
\begin{align}
 \Sigma_U^{ab}[\mathcal{G}^{ab}(\tau)]\! = \! -\rmi U \delta(\tau) \delta_{ab} \mathcal{G}^{ab}(\tau)\! -\! U^2 ab \left|\mathcal{G}^{ab}(\tau) \right|^2 \! \mathcal{G}^{ab}(\tau)\,, \label{eq:IPT}
\vspace{-0.5em}
\end{align}
This perturbative treatment of the impurity provides a good qualitative picture of the many-body physics of the lattice problem and has already been used out of equilibrium~\cite{neqDMFT}.
The whole procedure, (i) to (iii), is repeated until convergence is achieved.
\begin{figure}[t]
\centerline{
\includegraphics[width=3.2cm,angle=-90]{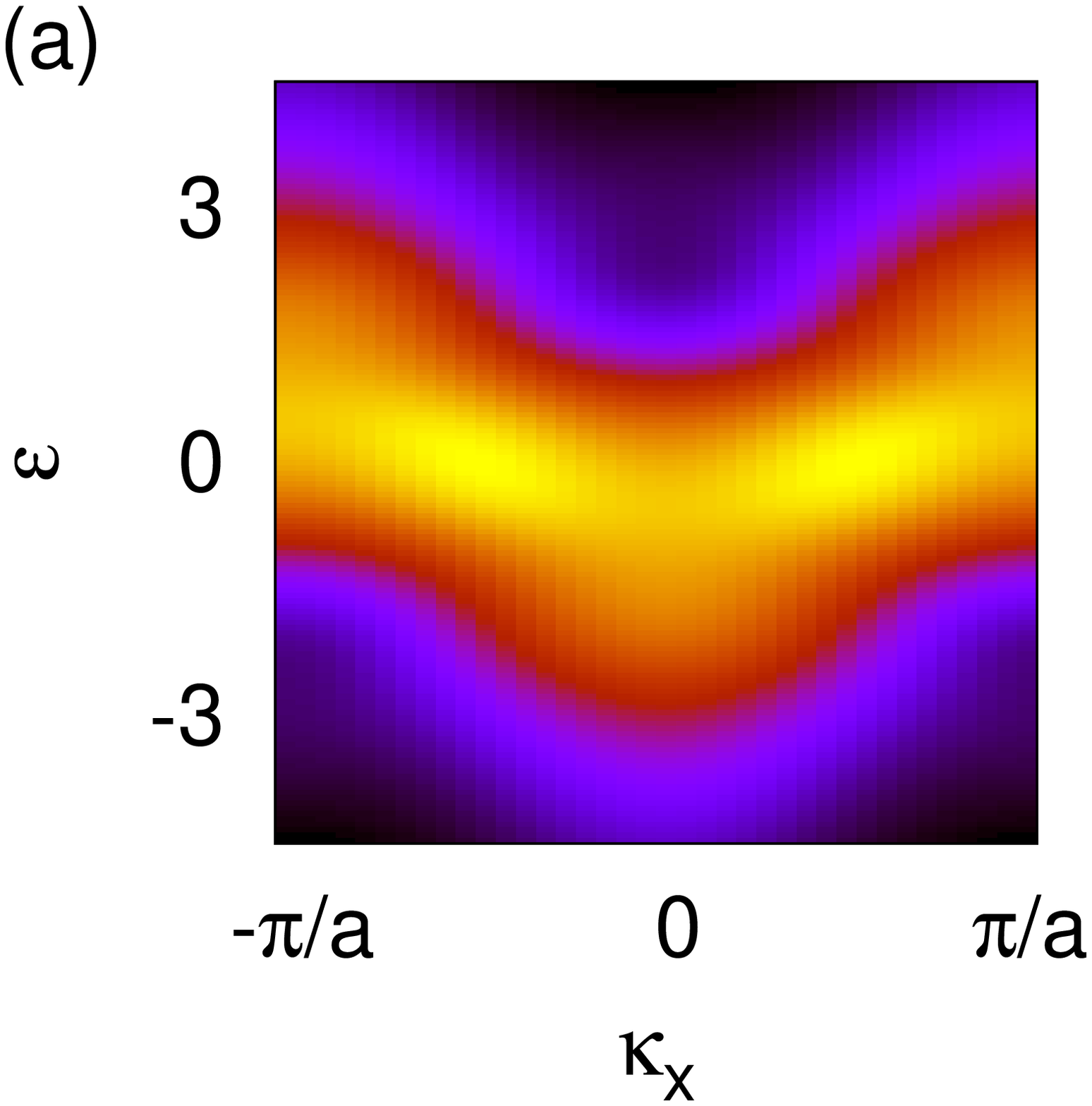}
\hspace{-2cm}
\includegraphics[width=3.2cm,angle=-90]{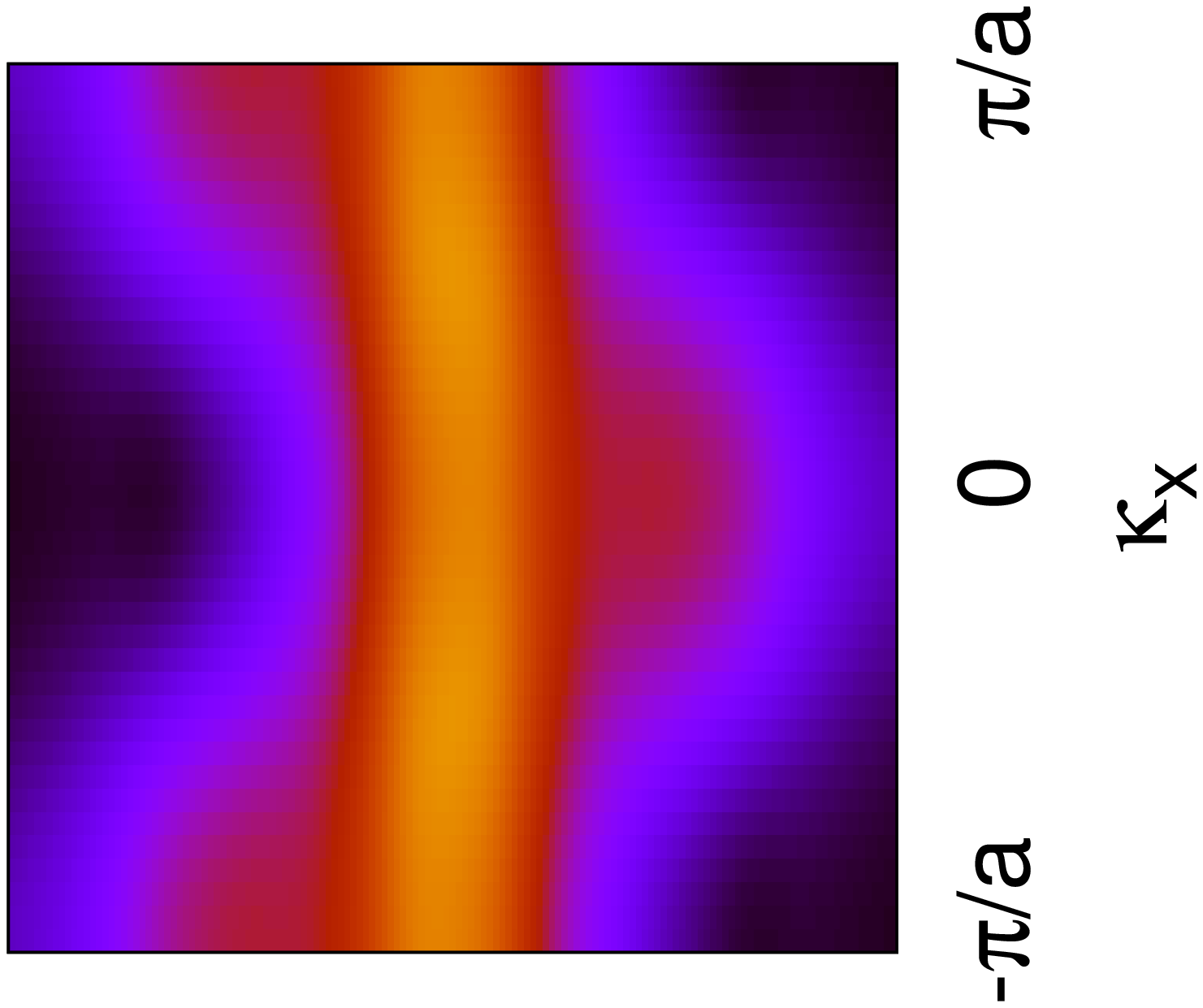}
\hspace{-1.9cm}
\includegraphics[width=3.2cm,angle=-90]{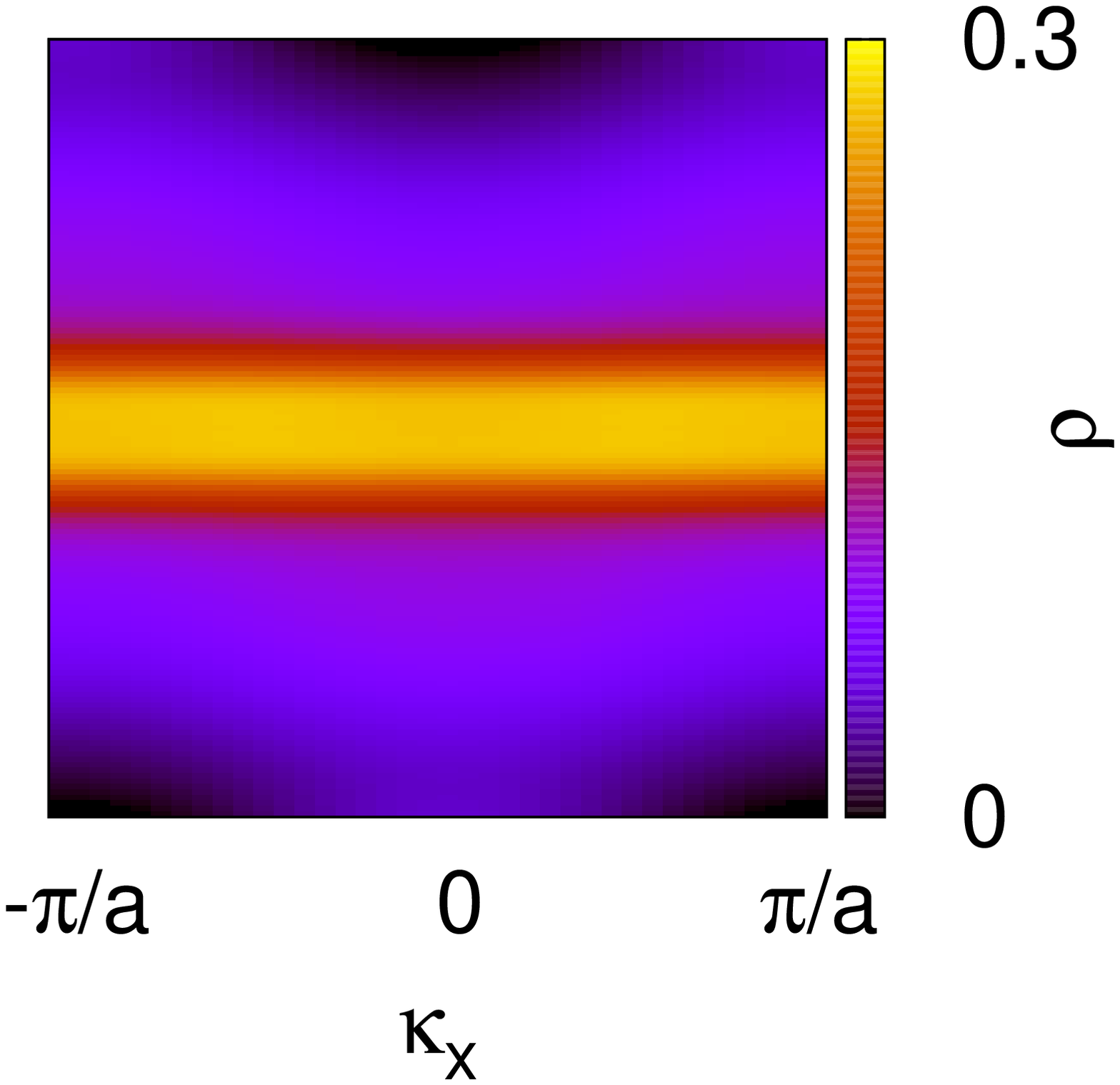}
}
\centerline{
\hspace{.1cm}
\includegraphics[width=3.2cm,angle=-90]{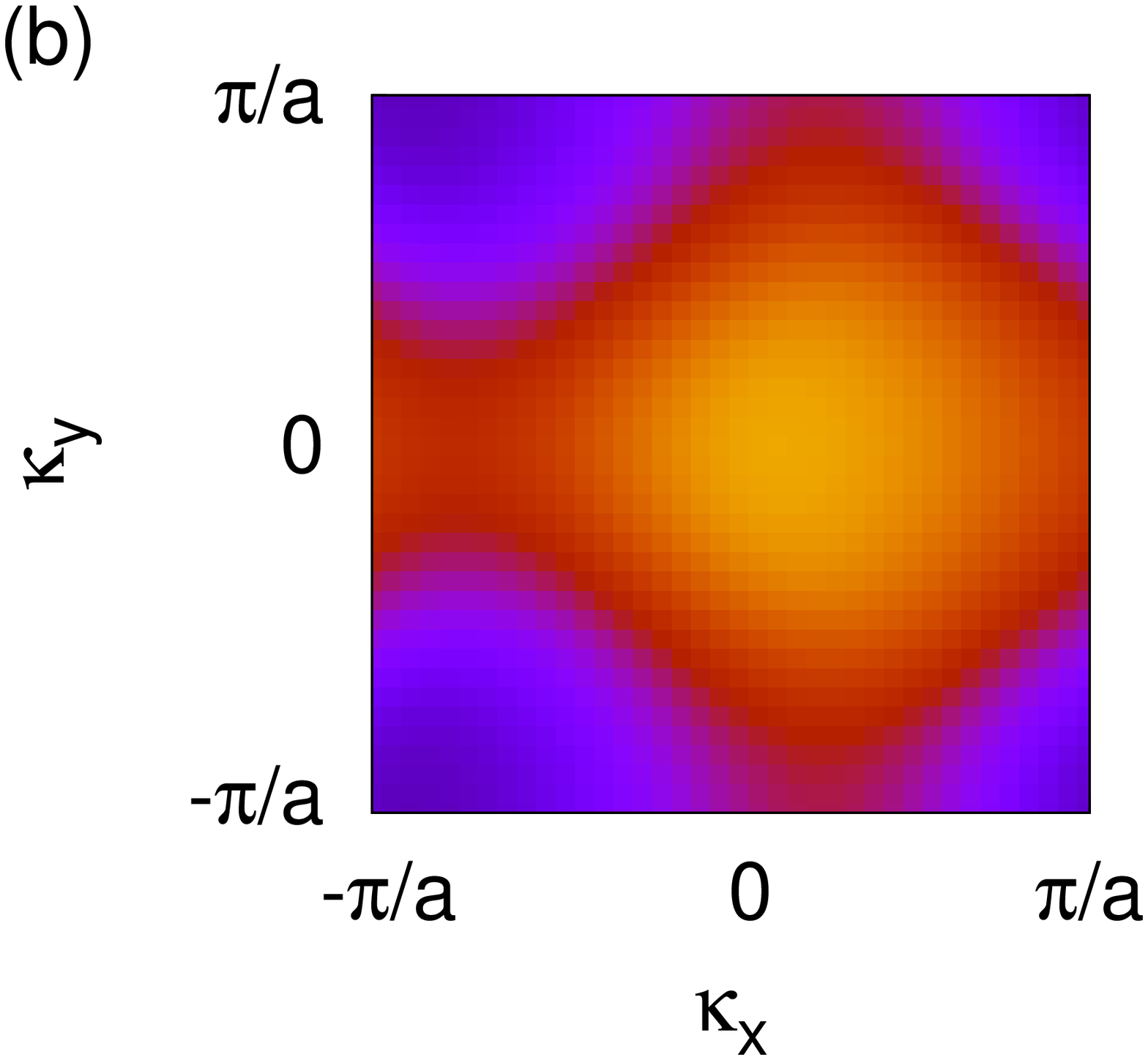}
\hspace{-2cm}
\includegraphics[width=3.2cm,angle=-90]{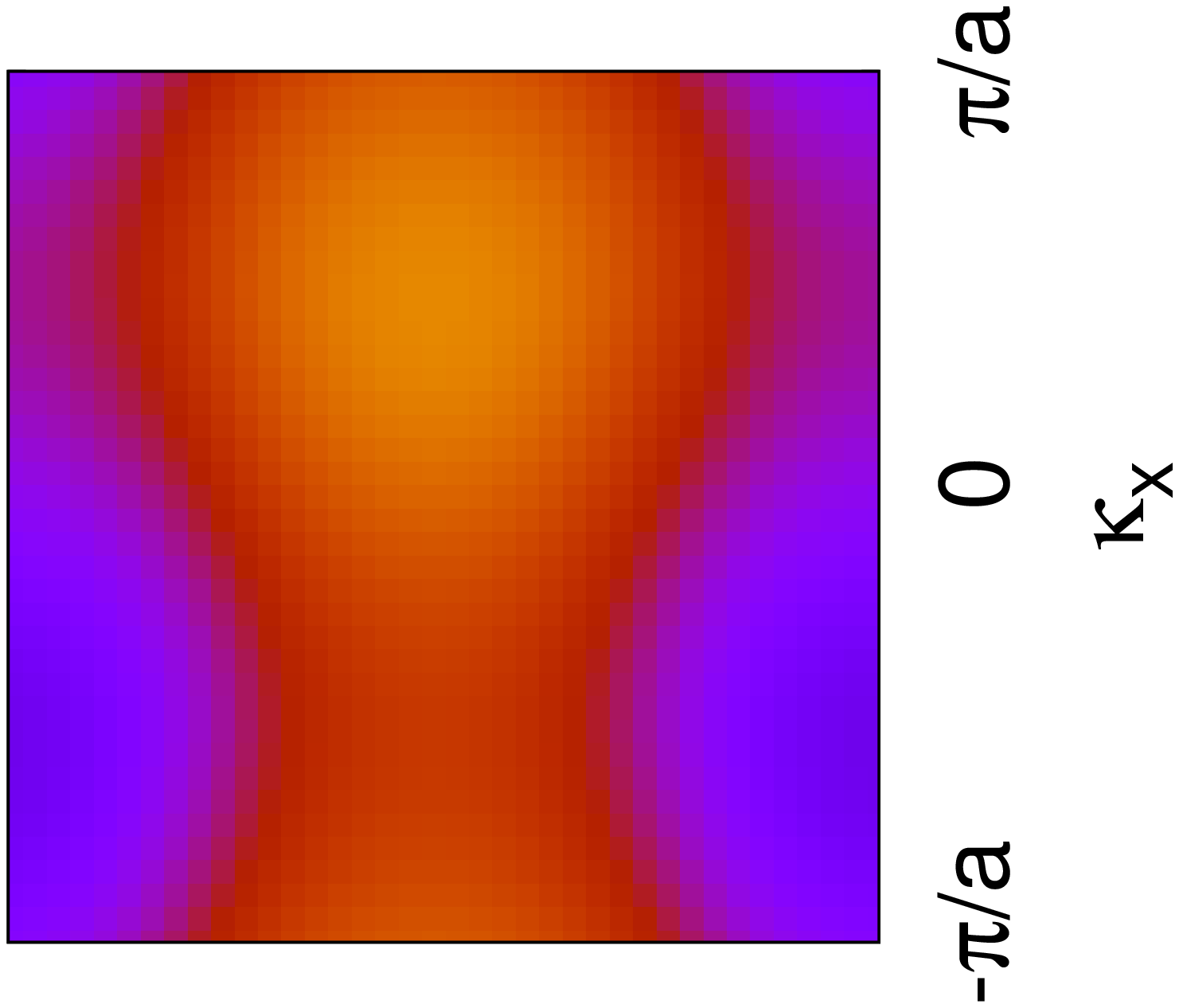}
\hspace{-2cm}
\includegraphics[width=3.2cm,angle=-90]{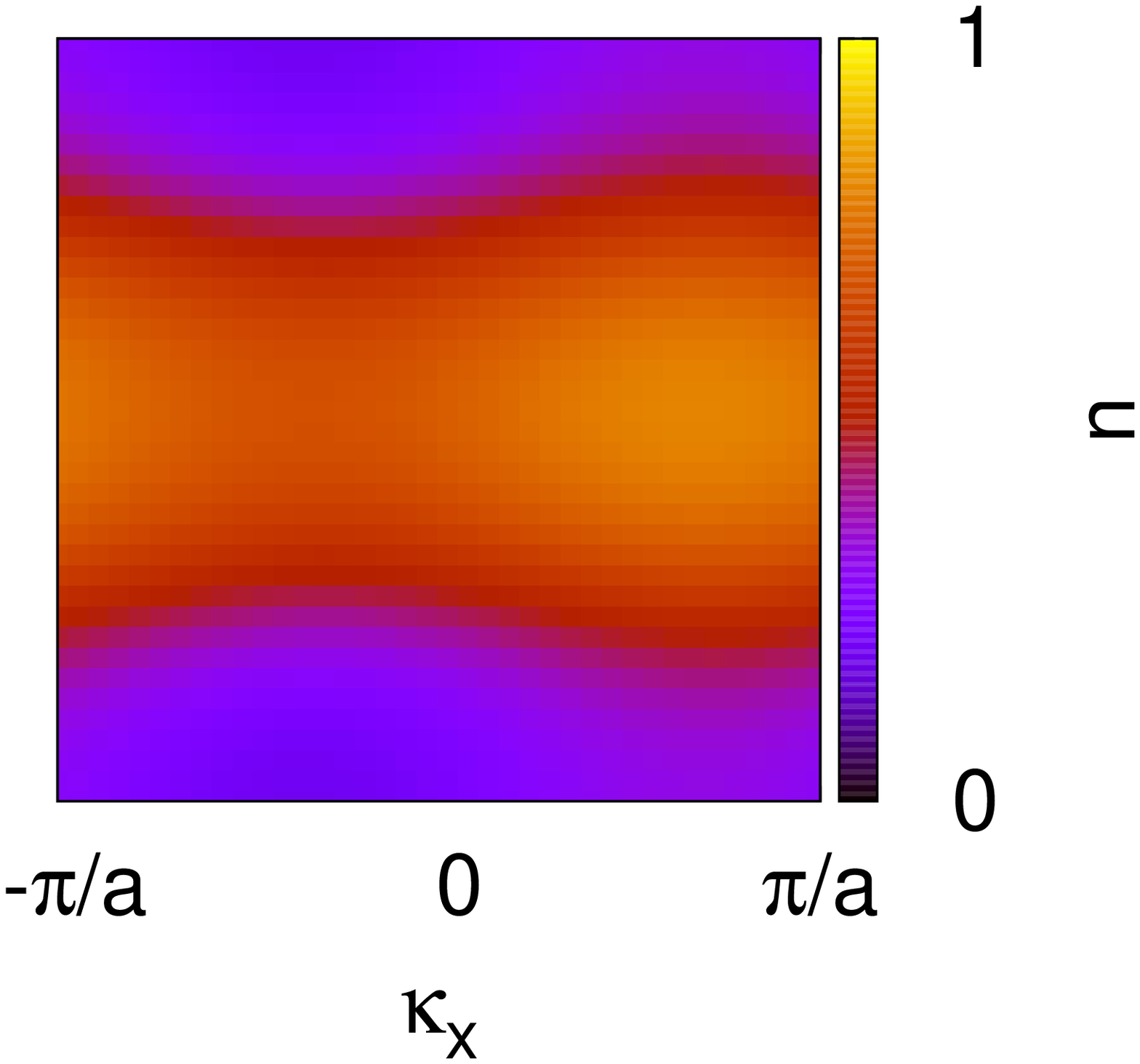}
}
\caption{\label{fig:photo} \footnotesize (color online)  (a)  Spectral function (integrated over $\kappa_y$) at $U=3$ for $E=3, 5$, and $10$.
(b) Momentum distribution function at $U=3$ for $E=1, 3$, and $5$.
 ($T\!=\!0.1, \Gamma\!=\!0.5$).}
\end{figure}

\paragraph{Results.---}We study the effect of the electric field and illustrate the dimensional crossover by implementing numerically the previous DMFT equations and by computing several physical observables as a function of the  field.
We consider the two-dimensional ($d=2$) Hubbard model on a square lattice with nearest-neighbor hoppings: $\epsilon(\boldsymbol{\kappa}) = \epsilon_0 \left[\cos(\kappa_x a)+\cos(\kappa_y a) \right]$. 
We work at half-filling and absorb the Hartree shift by expressing Green's functions in terms of the variable $\varpi'\equiv\varpi-U/2$. Below, we drop the prime.
The density of states of the thermal baths is chosen to be a Gaussian with a width controlled by $W$, yielding $\Sigma_{th}^R(\varpi) = -\rmi \Gamma w(\varpi/\sqrt{\pi}W)$ where $w(x) \equiv \left[1-\mathrm{erf}(-\rmi x) \right]\rme^{-x^2}$ is the Faddeeva function. $\Sigma^K_{th}$ can be expressed using the FDT (and $\mu_0=0$).
We concentrate on the effects of the electric field on the metallic phase.
We work at very small temperature $T$,  small dissipation $\Gamma$ (but large enough to allow for a stable steady state), and at fixed hopping amplitude $\epsilon_0$. Hereafter, all numerical results are obtained with $a=q=1$ and all energies are given in units of $\epsilon_0$.

\paragraph{Spectral and momentum distribution function.}
In Fig.~\ref{fig:photo}(a), we study the out-of-equilibrium spectral function $\rho(\epsilon,\boldsymbol{\kappa}) =- \mbox{Im}\,G^R(\kappa)|_{\varpi=\epsilon} /\pi$   averaged over $\kappa_y$. This quantity can be accessed experimentally by angle resolved photoemission spectroscopy (ARPES).
In Fig.~\ref{fig:photo}(b), we also plot the momentum distribution function $n(\boldsymbol{\kappa}) = 1/2 - G^K(\tau\!=\!0;\boldsymbol{\kappa})$.
To illustrate the crossover from the two-dimensional to the one-dimensional system, we show that both observables lose their dependence on $\kappa_x$ as the field intensity is increased, until they are completely invariant under $\kappa_x$ translations. 

\begin{figure}[t]
\centerline{
\includegraphics[width=3.4cm,angle=-90]{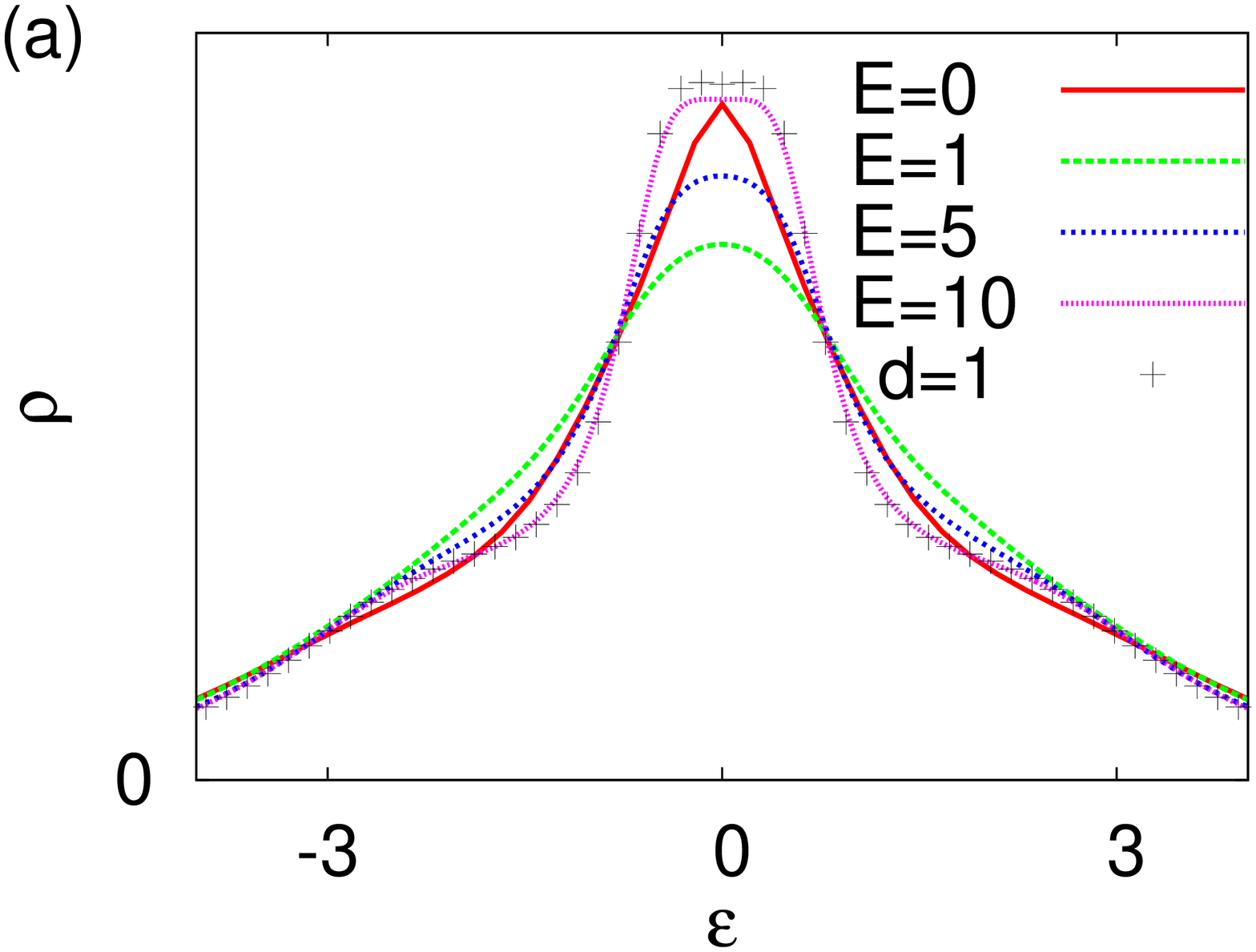}
\hspace{-0.7cm}
\includegraphics[width=3.4cm,angle=-90]{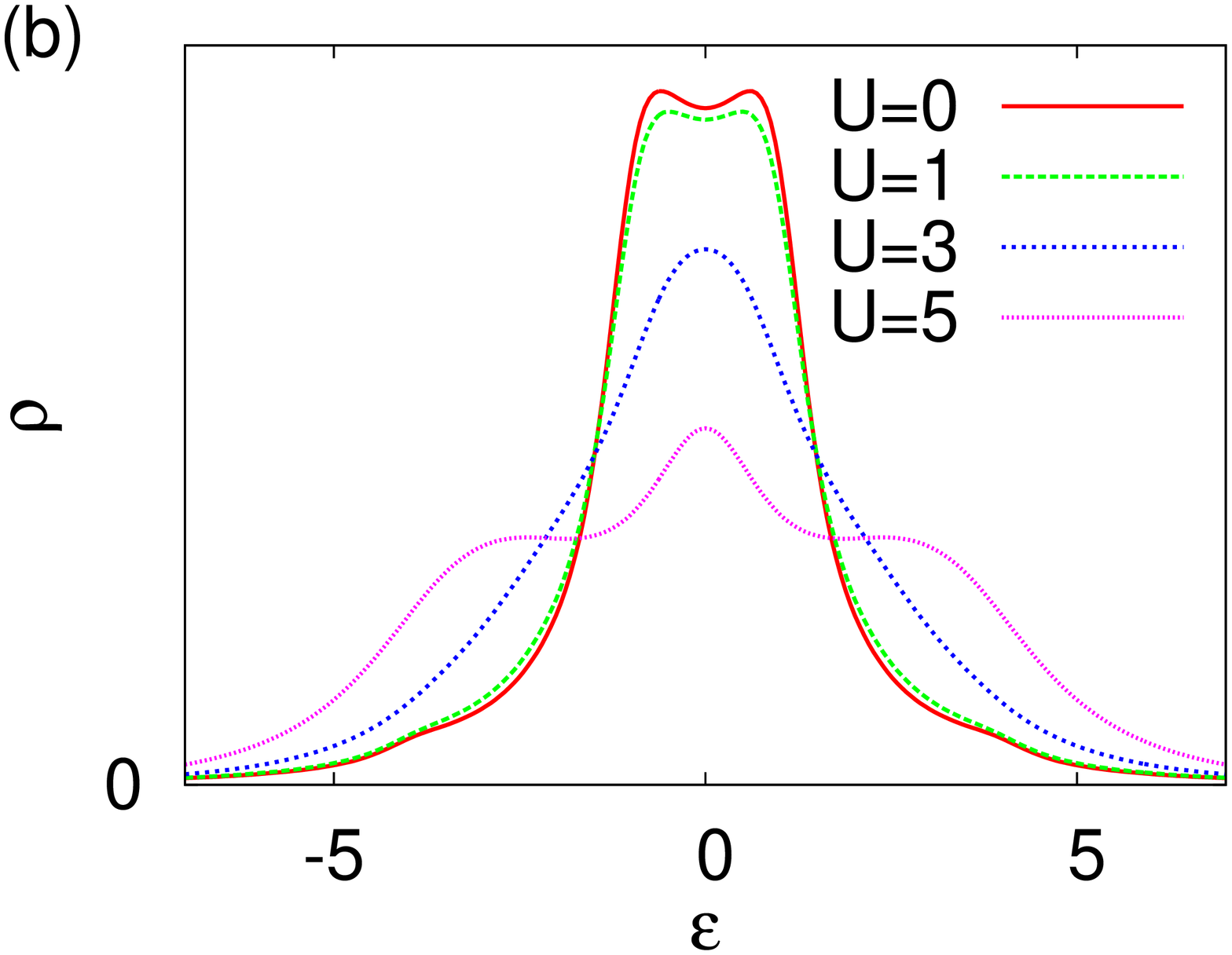}
}
\caption{\label{fig:DOS} \footnotesize (color online)  (a)  Density of states at fixed repulsion ($U=3$) for different values of the electric field. The crosses correspond to the equilibrium $d=1$ Hubbard model. (b) Density of states at fixed field ($E=3$) for different values of the electronic repulsion. ($T\!=\!0.1, \Gamma\!=\!0.5$).}
\end{figure}

\paragraph{Density of states.}
For a fixed $U$, the effect of the electric field on the local density of states (DOS) $\rho(\epsilon) = -\mbox{Im}\, G_\mathrm{}^R(\varpi=\epsilon)/\pi$ is illustrated in Fig.~\ref{fig:DOS}(a). Starting from the equilibrium metallic solution, the quasi-particle peak is first rounded up for intermediate field intensities. For very strong fields, it develops a structure peculiar to the equilibrium paramagnetic DMFT solution of the $d=1$ Hubbard model (the DOS of which is given with crosses for comparison).
In Fig.~\ref{fig:DOS}(b) we follow the local density of states when varying $U$ at fixed $E$. For $0 = U\ll |q|Ea$, the double peak structure of the (non-interacting) $d=1$ Hubbard model is recovered. For increasing values of $U$, the onset of the Mott transition occurs as the quasi-particle peak decreases and the insulating lobes grow. 
\begin{figure}[t]
\centerline{
\hspace{-.1cm}
\includegraphics[width=3.4cm,angle=-90]{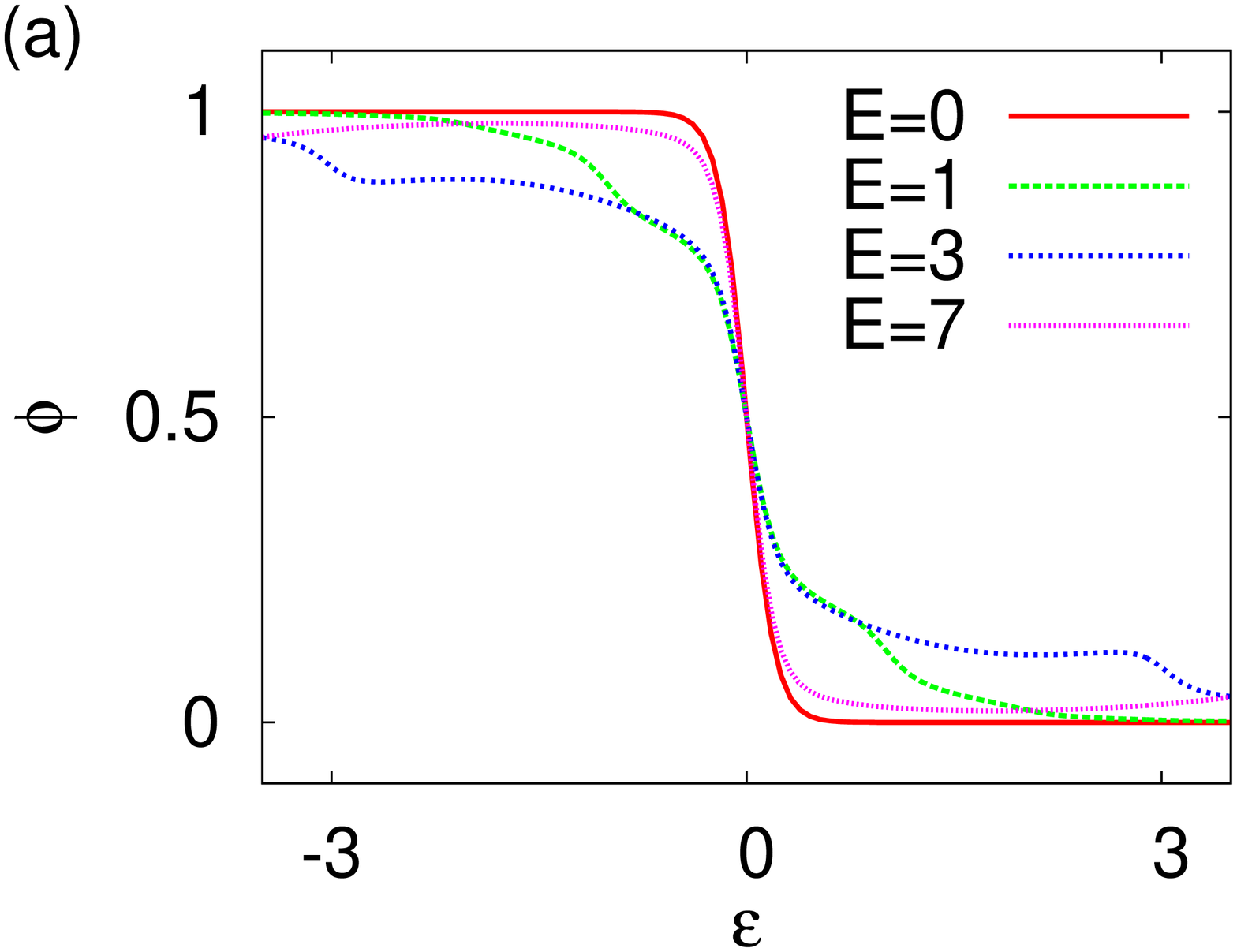}
\hspace{-.75cm}
\includegraphics[width=3.4cm,angle=-90]{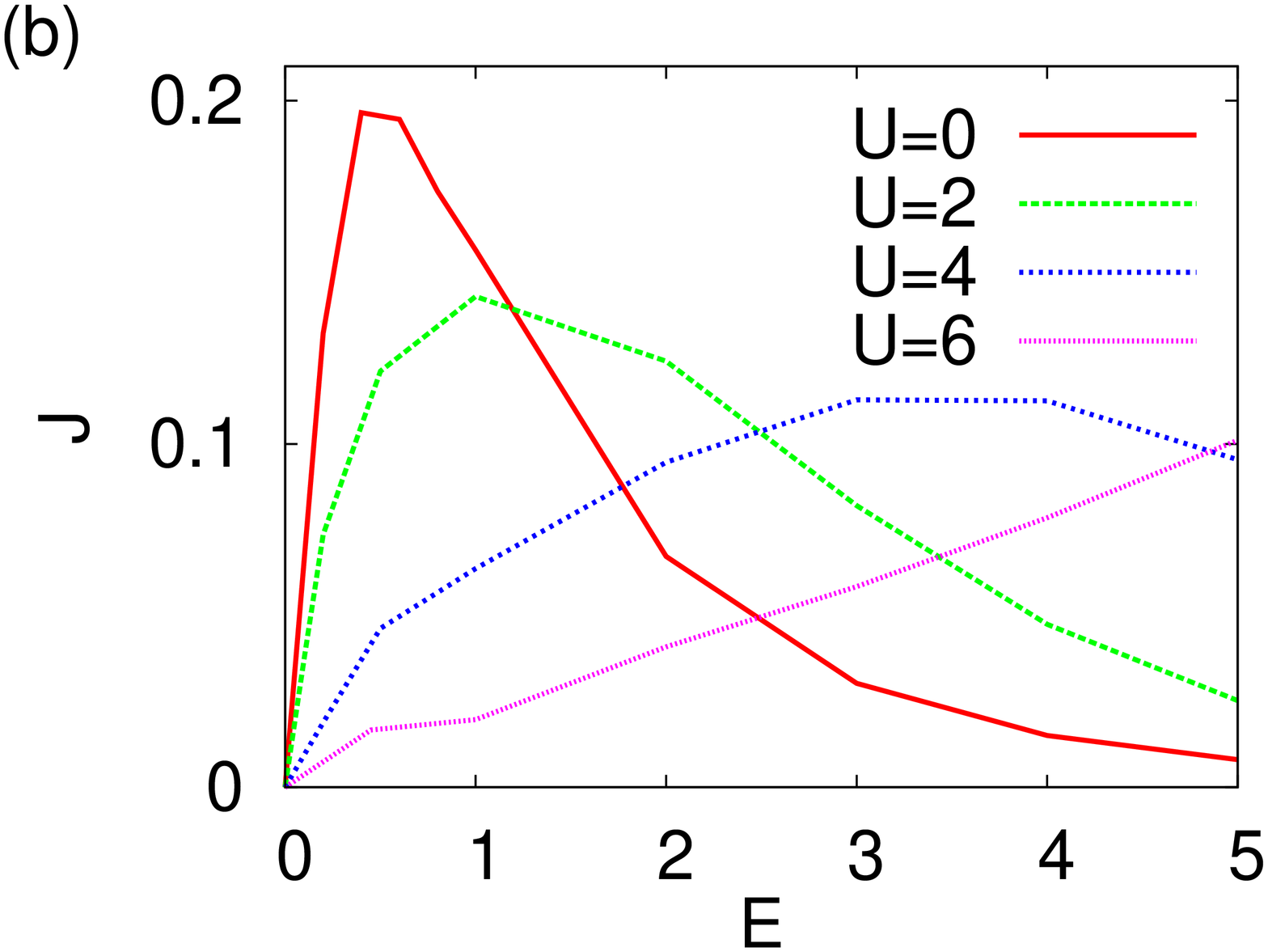}
}
\caption{\label{fig:current} \footnotesize (color online) (a) Distribution function for different values of the electric field at fixed $U=3$. Notice how the distribution function starts from a Fermi-Dirac distribution for $E=0$, and how it ends as a Fermi-Dirac distribution for $E=7$ ($T\!=\!0.1, \Gamma\!=\!0.5$). (b) Steady-state current density as a function of the electric field for different values of the electronic repulsion ($T\!=\!0.05, \Gamma\!=\!0.25$).}
\end{figure}

\paragraph{Distribution function.} 
In equilibrium, the FDT implies that the distribution function is $\boldsymbol{\kappa}$-independent and given by the Fermi-Dirac distribution  $f(\epsilon)\equiv [1+\rme^{({\epsilon-\mu_0})/{T}} ]^{-1}$. Out of equilibrium, it is now $\boldsymbol{\kappa}$-dependent and we study in Fig.~\ref{fig:current}(a) the local energy distribution, corresponding to the distribution function of the impurity, $\phi(\epsilon) \equiv \frac{1}{2} \left[1+G^K_\mathrm{}(\varpi)/ \mbox{Im}\, G_\mathrm{}^R(\varpi) \right]_{\varpi=\epsilon}$. For weak fields, electrons located just below the Fermi level are excited to states above the Fermi level. This can be seen as the system heating up, increasing the heat transfer to the reservoirs and balancing the injected power~\cite{AronKotliar2}.
As the field intensity is increased, this pocket of electrons sent above the Fermi surface gets broader and the distribution function shows two steps at energies of order $\pm qEa$.
These steps get wider but also thinner for a very strong field and the distribution approaches the Fermi-Dirac one, showing that equilibrium is reached.

\paragraph{Current.}
In Fig.~\ref{fig:current}(b) we study the steady-state current density, $\mathbf{J}\propto 2 q \int \ud{\kappa} G^K(\kappa) \boldsymbol{\nabla}_{\!\boldsymbol{\kappa}} \epsilon(\boldsymbol{\kappa})$, as a function of the field intensity and $U$. 
The establishment of a finite current is made possible by two scattering processes.
One is due to the electronic interaction whereas another is due to the coupling to the dissipative bath [see \textit{e.g.} the finite current at $U=0$ in Fig.~\ref{fig:current}(b)].
For all values of the electronic interaction, the steady-state current is first linear with the field intensity with a linear conductivity which is a decreasing function of $U$, as expected by linear response theory.
After going through a maximum, the current is exponentially damped at very strong fields. 
The damping is again consistent with the dimensional reduction occurring at very strong fields: the physics are the ones of an equilibrium (current-less) Hubbard model in one dimension.
A similar non-linear characteristic has already been found in the $2d$ $t-J$ model~\cite{Vidmar} where the spin degrees of freedom were claimed to act as a dissipative background. For the case of a Hubbard sample strapped between two leads at different chemical potentials, a maximum in the current-voltage characteristic was also found in the case of a two-leg ladder~\cite{Knap} but not in the case of a $3d$ sample~\cite{Okamoto}, maybe due to a lack of a dissipative mechanism in the bulk.

\paragraph{Conclusion.---}
We have investigated a correlated metal in a non-equilibrium steady-state regime driven by a constant electric field.
By taking into account the need for a  heat dissipation mechanism, and by writing the Schwinger-Dyson equations in a covariant fashion, we revealed a crossover to a lower dimensional system in equilibrium as the field intensity is increased.
We have generalized the equilibrium DMFT equations to the case of correlated systems driven by electric field, which allows a comprehensive understanding of the steady-state solution at a relative low computational cost.
By solving the impurity in the IPT approximation, we illustrated the dimensional crossover on a two-dimensional system. 
We showed that $\boldsymbol{\kappa}$-dependent observables lose their dependence in the direction of the field as this one is increased. 
We repeated the analysis for local quantities and showed their convergence to equilibrium.
In particular, the distribution function was shown to converge towards the Fermi-Dirac distribution, completing the proof that 
it is indeed equilibrium which is reached in the $d-1$ dimensions.

We thank Adriano Amaricci,  Olivier Parcollet, and Youhei Yamaji for insightful discussions.
This work has been supported by NSF DMR-0906943. C. W. was supported by the Swiss National Foundation for Science.

\end{document}